\documentclass[runningheads]{llncs}
\usepackage[T1]{fontenc}
\usepackage{graphicx,url}
\usepackage{amsmath}
\usepackage{amssymb}
\usepackage{booktabs}
\usepackage{algorithm}
\usepackage{algpseudocode}
\usepackage{xcolor}
\usepackage{multirow}
\usepackage{color}

\begin{document}
\title{Tracing and Taming Hallucinated Terms via Subsequence-Association Filtering and Breadth-Aware Selection for LLM Query Expansion}
%
\titlerunning{SAFE-QE: Taming Hallucinated Query Expansion}
\author{Amit Kumar Jaiswal\inst{1} \and
Sagar Uprety\inst{2} \and
Prayag Tiwari\inst{3}}
\authorrunning{A. K. Jaiswal et al.}
%
\institute{Indian Institute of Technology (BHU), Varanasi, India \\
\email{amit.chr@iitbhu.ac.in} \and
University College London, London, UK \\
\email{s.uprety@ucl.ac.uk} \and
Halmstad University, Halmstad, Sweden \\
\email{prayag.tiwari@hh.se}}
\maketitle              
\begin{abstract}
Large language models (LLMs) are now a standard source of query expansion terms but they hallucinate. They inject non existent entities and off-topic terms that add ranking noise and can hurt strong retrievers. Existing remedies such as corpus grounding, mutual verification and preference alignment treat hallucinated terms as an undifferentiated nuisance to be filtered by agreement
heuristics, without a mechanistic account of why a given term appears or a principled criterion for how many terms to keep. We propose Subsequence-Association Filtered, breadth-aware query expansion (SAFE-QE), a training-free, retriever agnostic method that draws on the subsequence-association account of hallucination. We score each candidate expansion term by its corpus conditioned cross-context reproducibility, terms
that reappear under randomised query contexts yet lack support in the pseudo-relevant set are flagged as products of dominant hallucinatory associations and down-weighted. Drawing on the characterisation of language generation as a trade-off between consistency (no hallucination) and breadth (coverage), we cast term selection as constrained coverage maximisation over the pseudo-relevant set, keeping a set that is approximately broad but consistent and stable across resampling. We also introduce the Hallucinated Expansion Term Rate, a query side faithfulness metric that counts injected expansion terms no relevant document supports. We evaluate our approach on benchmarked TREC and BEIR collections. Our SAFE-QE is statistically indistinguishable from generative and corpus-steered expansion on nDCG@10 while injecting fewer unsupported terms 4.3 per query against 27.8 for Query2doc and 57.5 for HyDE and on a strong learned sparse retriever, where every expansion method regresses. Faithfulness is therefore available at no measurable cost in ranking quality.

\end{abstract}
\section{Introduction}\label{sec:intro}
Query expansion augmenting a short query with related terms to close the vocabulary gap between queries and documents is one of the oldest and most reliable levers in information retrieval (IR). The current wave of query expansion replaces corpus derived pseudo-relevance feedback (PRF) with terms, keywords, or hypothetical documents generated by large language models (LLMs). Methods such as Query2doc~\cite{wang2023query2doc}, HyDE~\cite{gao2023precise}, query expansion
by prompting~\cite{jagerman2023query}, and generative relevance feedback~\cite{mackie2023generative} report substantial gains, especially for lexical
retrievers such as BM25. Yet these gains are fragile. LLMs hallucinate, they produce fluent text that is factually wrong or unsupported by any relevant document. In query expansion, this manifests as injected non existent entities and off-topic terms. Two empirical studies make the cost concrete. Weller et al.~\cite{weller2024generative} show
across many methods, retrievers, and datasets, a negative correlation between retriever strength and the benefit of expansion, generated expansions help weak retrievers but add additional noise that harms strong ones by introducing false positives near the top of the ranking. Abe et
al.~\cite{abe2025llm} identify two failure modes unfamiliar queries, where the LLM introduces non existent entities or unrelated terms, and ambiguous queries, where expansions collapse onto the most popular sense and narrow coverage. Recent work to ground generation in the corpus are CSQE~\cite{lei2024corpus}, LameR~\cite{shen2023large},
ThinkQE~\cite{lei2025thinkqe} to cross verify generated and retrieved
evidence~\cite{jia2024mill}, to filter smaller model generations~\cite{liu2025golfer}, or to align the generator with ranking
preferences via training~\cite{yao2025llm}. These are effective but they share two blind spots. They are mechanism agnostic, a term is
kept or dropped by whether it happens to co-occur with retrieved evidence. They lack a principled selection criterion, how many terms to keep, and how to trade coverage against faithfulness is left to a fixed top-$k$ or a tuned weight.
\begin{figure}
    \centering
    \includegraphics[width=\linewidth]{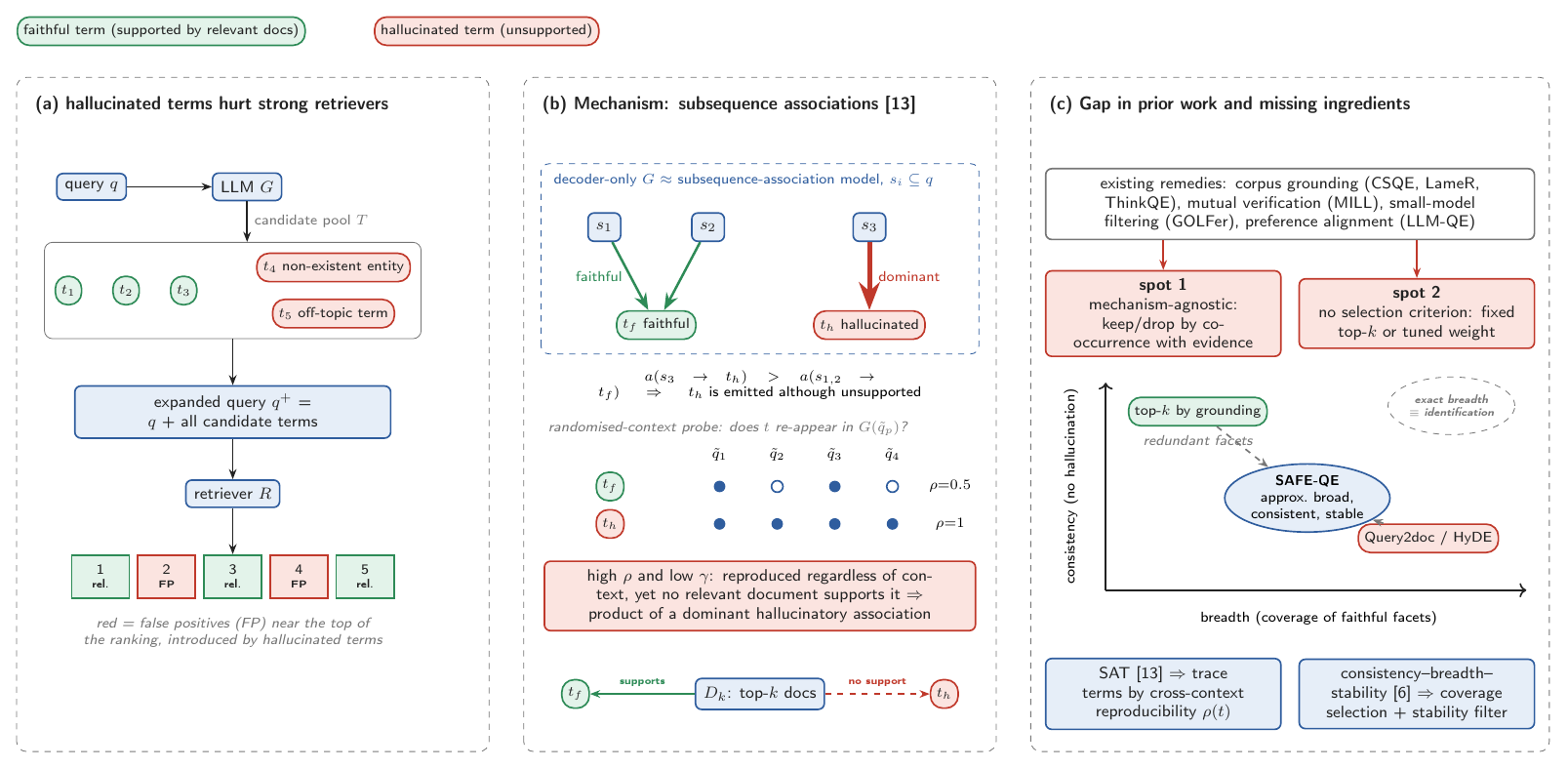}
    \caption{A schematic pipeline of a dominant hallucinated subsequence association makes the LLM emit context-free, ungrounded expansion terms that inject false positives into the ranking, and these terms hurt stronger retrievers.}
    \label{fig:intuition}
\end{figure}
We argue that two recent theoretical results~\cite{sun2026and,kalavasis2026characterizations} supply exactly the missing ingredients, and that is understudied in IR. Sun et
al.~\cite{sun2026and} explain hallucination mechanistically, decoder only transformers behave as subsequence-association models, and hallucination occurs when a dominant hallucinatory association between an input subsequence and an output token outweighs faithful ones. Also, Kalavasis et al.~\cite{kalavasis2026characterizations} characterise language generation as
an intrinsic trade-off between consistency i.e., eventually producing only
valid outputs or no hallucination, and breadth refers to covering a rich subset of the target, formalising exact and approximate breadth and the role of generator stability. We detail our underlying notion schematically in Fig.~\ref{fig:intuition}.

Inspired by these approaches, we propose SAFE-QE that treats expansion term generation as generation-in-the-limit over the target corpus vocabulary and asks for a term set that is approximately broad but
consistent, broad enough to improve recall, yet consistent enough to avoid injecting hallucinated terms. It (a) generates candidate terms with an open LLM,
(b) scores each by a lightweight, corpus conditioned analogue of SAT that
rewards terms grounded in the pseudo-relevant set and penalises terms whose
presence is reproducible across randomised contexts but corpus unsupported,
(c) selects a set by submodular coverage maximisation under a grounding
constraint. And, (d) keeps only terms that are stable across resampled
generations. SAFE-QE is training-free and retrieve agnostic wrapping any
generate-then-expand pipeline.\\
Our contributions can be summarised as follows.
\begin{enumerate}
  \item We reframe hallucination in query expansion through the lens of subsequence
  associations~\cite{sun2026and} and the consistency breadth stability
  characterisation of language generation~\cite{kalavasis2026characterizations}.
  \item We introduce \textbf{SAFE-QE}, a training-free method combining a
  corpus conditioned reproducibility score with breadth-aware, stability checked
  selection.
  \item We propose \textbf{HETR}, a Hallucinated Expansion Term Rate that quantifies unsupported injected terms via atomic corpus verification, adapting the HALoGEN method~\cite{ravichander2025halogen} to query expansion.
\end{enumerate}

\section{Related Work}
\textbf{LLM based query expansion:} LLM QE methods cluster into three families. Pseudo document generation appends an LLM written document to the query, Query2doc~\cite{wang2023query2doc} concatenates a generated passage (with query repetition for BM25) and reports 3-15\% BM25 gains on MS~MARCO and TREC~DL. HyDE~\cite{gao2023precise} encodes a
zero shot `hypothetical document' with an unsupervised dense encoder whose
bottleneck is credited with filtering hallucinations. MuGI~\cite{zhang2024exploring}
fuses multiple pseudo references with adaptive reweighting. Term or keyword generation prompts for expansion terms with
chain of thought giving the strongest keywords~\cite{jagerman2023query}. Corpus grounded or feedback fused methods reduce hallucination by conditioning on retrieved evidence. GRF~\cite{mackie2023generative} builds RM3 based feedback models from long form LLM text. CSQE~\cite{lei2024corpus} extracts
pivotal sentences from BM25's top-$k$. LameR~\cite{shen2023large} conditions
generation on retrieved documents. ThinkQE~\cite{lei2025thinkqe} iterates a
thinking process against the corpus. On the document side, Doc2Query~\cite{gospodinov2023doc2query} filters hallucinated generated queries before indexing, the closest prior instance of filtering generated text, but applied to documents, heuristically, and without a
generative mechanism account.\\
\\
\textbf{Hallucination in expansion, and why it matters}
Weller et al.~\cite{weller2024generative} and Abe et al.~\cite{abe2025llm} document that generated expansions inject noise and non existent entities, hurting strong retrievers and unfamiliar or ambiguous queries. Yoon et
al.~\cite{yoon2025hypothetical} argue that much of LLM-QE's benefit stems from knowledge leakage, gains appear mainly when generated text contains gold entailed content. This is conceptually adjacent to our formulation, both links to memorised associations but~\cite{yoon2025hypothetical} diagnose leakage as a confound, whereas we exploit the same mechanism to separate faithful from
hallucinatory associations and filter accordingly. Mitigations to date rely on
grounding i.e. CSQE, LameR, ThinkQE, agreement such as MILL~\cite{jia2024mill}, small model filtering~\cite{liu2025golfer}, or training LLM-QE~\cite{yao2025llm}.
None grounds mitigation in a theory of the generative mechanism, and none frames
selection as a consistency breadth trade-off.

Decoder-only transformers act as subsequence embedders and that hallucination arises when a dominant hallucinatory subsequence association outweighs faithful ones~\cite{sun2026and}. Their SAT algorithm identifies causal input subsequences by measuring how reproducibly they trigger a target output across randomised contexts, validated on HALoGEN~\cite{ravichander2025halogen}. Kalavasis et al.~\cite{kalavasis2026characterizations}, extends on generation in the limit, prove that generating with exact breadth is as hard as identification, that approximate breadth (missing only finitely many valid items) is strictly easier, and that requiring stability (a generator whose output
eventually stops changing) collapses these notions to the harder regime. A sharp consistency breadth stability interplay. We are to our knowledge, the first to port either result into information retrieval.
\section{Our Method}

\subsection{Problem Formulation}
Given a query $q$ and a corpus $\mathcal{C}$ with a first-stage retriever $R$,
query expansion produces an expanded query $q^{+}$ that retrieves a better ranking. An LLM
$G$ proposes a pool of candidate terms $\mathcal{T}=\{t_1,\dots,t_m\}$ (tokens,
phrases, or salient spans of a generated pseudo-document). Some are
faithful supported by documents relevant to $q$ and some are hallucinated, fluent but unsupported. Our goal is to keep a subset
$\mathcal{T}^\star\subseteq\mathcal{T}$ that maximises coverage of the relevant
material (breadth) while excluding hallucinated terms (consistency).

We view $G$ as a generator over the corpus vocabulary and target the approximately broad but consistent regime of
\cite{kalavasis2026characterizations}. We do not demand every valid term, exact
breadth is as hard as identification and needlessly risky, only that
$\mathcal{T}^\star$ miss few faithful facets while admitting no hallucinated ones. To decide faithfulness per term, we use a corpus conditioned analogue of the subsequence association signal of~\cite{sun2026and}. 

\subsection{Candidate generation}
We prompt $G$, an open instruction-tuned LLM with a Query2doc/CoT-style template to produce $N$ independent samples, yielding a multiset of candidate terms with sample frequencies. Let $\mathrm{freq}(t)\in[0,1]$ denote the
fraction of the $N$ samples containing $t$. High $\mathrm{freq}$ is our practical proxy for a \emph{stable} generator in the sense
of~\cite{kalavasis2026characterizations}, a term the generator commits to under
resampling.

\subsection{Corpus conditioned reproducibility score}
We compute two quantities for each candidate $t$.
\paragraph{Cross context reproducibility $\rho(t)$.} Following the SAT approach~\cite{sun2026and}, we build a small perturbation set of $P$ variant prompts that randomly drop and recombine subsequences of $q$ (and, when used, of the generated pseudo-document), then measure how often $t$ still appears,
\begin{equation}
  \rho(t) = \frac{1}{P}\sum_{p=1}^{P}\mathbf{1}\!\left[\,t \in G(\tilde q_p)\,\right].
\end{equation}
A high $\rho(t)$ means $t$ is triggered by a dominant association that is largely independent of the specific query intent, a hallmark of a context-insensitive (often hallucinatory) association.

\paragraph{Corpus grounding $\gamma(t)$.} Let $D_k$ be the top-$k$ pseudo-relevant documents from $R(q)$. We set $\gamma(t) = \max_{d \in D_k}\, \mathrm{sup}(t, d)$, where $\mathrm{sup}(t,d)\in[0,1]$ is a support score, lexical presence for sparse settings, or maximum cosine similarity between the embedding of $t$ and the sentences of $d$ for dense settings.

\paragraph{Faithfulness.} A term is trusted when it is \emph{grounded}, and it is \emph{suspected hallucinatory} when it is highly reproducible yet ungrounded. We combine these into a keep-score
\begin{equation}\label{eq:score}
  s(t) = \underbrace{\gamma(t)}_{\text{grounding}}
            - \lambda \,\underbrace{\rho(t)\,\bigl(1-\gamma(t)\bigr)}_{\text{context-free \& ungrounded}}
            + \mu\,\mathrm{freq}(t),
\end{equation}
with non-negative hyperparameters $\lambda,\mu$. The middle term is of essence, it penalises exactly the terms flagged by the subsequence-association account, those a dominant association reproduces regardless of context and that no relevant document supports.

\subsection{Breadth-aware, consistency constrained selection}
Keeping the top-$k$ by $s(t)$ ignores redundancy, many high-scoring terms may
cover the same facet. We instead maximise coverage of the pseudo-relevant
set subject to a consistency (grounding) constraint. We represent each document
$d\in D_k$ by the set of candidate terms it supports,
$\mathrm{cov}(t)=\{d\in D_k : \mathrm{sup}(t,d)\ge\tau\}$. We solve
\begin{equation}\label{eq:submod}
  \mathcal{T}^\star = \arg\max_{\substack{S\subseteq \mathcal{T}_g,\;|S|\le B}}
  \; \Bigl| \bigcup_{t\in S}\mathrm{cov}(t)\Bigr|
  + \beta \!\sum_{t\in S} s(t),
  \quad
  \mathcal{T}_g=\{t:\gamma(t)\ge\tau_g\},
\end{equation}
i.e., a budget-$B$ maximum coverage objective which is submodular, solved greedily with
a $(1-1/e)$ guarantee restricted to grounded candidates $\mathcal{T}_g$. The
coverage term encodes \emph{approximate breadth} span the facets present in
the corpus, while the grounding constraint enforces \emph{consistency}. This
is the operational reading of the trade-off
in~\cite{kalavasis2026characterizations}, we deliberately forgo exact breadth (all valid terms) for a broad but safe subset.

\subsection{Stability filter and expansion}
We drop any $t\in\mathcal{T}^\star$ with $\mathrm{freq}(t)$ below a threshold
$\theta$, volatile terms that appear in few resamples are empirically more
often hallucinated. The surviving terms form $q^{+}$, for sparse retrieval we
fuse them into an RM3-style expansion with weight $\alpha$, for dense retrieval
we append them to $q$ before encoding. Our algorithm~\ref{alg:safeqe} summarises the
procedure.

\begin{algorithm}[t]
\caption{SAFE-QE}\label{alg:safeqe}
\begin{algorithmic}[1]
\Require query $q$, retriever $R$, LLM $G$, budget $B$, params $\lambda,\mu,\beta,\tau,\tau_g,\theta,\alpha$
\State $\{$samples$\}\gets G(q)$ sampled $N$ times; form candidate pool $\mathcal{T}$ with $\mathrm{freq}(\cdot)$
\State $D_k \gets$ top-$k$ documents of $R(q)$ \Comment{pseudo-relevant set}
\For{each $t\in\mathcal{T}$}
  \State compute $\rho(t)$ over $P$ perturbed prompts; $\gamma(t)$ over $D_k$
  \State $s(t)\gets \gamma(t)-\lambda\,\rho(t)(1-\gamma(t))+\mu\,\mathrm{freq}(t)$ \Comment{Eq.~\eqref{eq:score}}
\EndFor
\State $\mathcal{T}_g\gets\{t:\gamma(t)\ge\tau_g\}$; solve Eq.~\eqref{eq:submod} greedily $\to \mathcal{T}^\star$
\State $\mathcal{T}^\star\gets\{t\in\mathcal{T}^\star:\mathrm{freq}(t)\ge\theta\}$ \Comment{stability filter}
\State \Return $q^{+}\gets \textsc{Fuse}(q,\mathcal{T}^\star,\alpha)$
\end{algorithmic}
\end{algorithm}
\section{Hallucination Metric for Query Expansion}\label{sec:metric}
Retrieval metrics measure whether expansion helps ranking, instead of whether it is
\emph{faithful}. A hallucinated term can occasionally help by lucky lexical
overlap. To measure faithfulness, we adapt the atomic verification protocol of HALoGEN~\cite{ravichander2025halogen} to query expansion. For a query $q$ with
judged relevant documents $\mathrm{Rel}(q)$, and a set of injected expansion
terms $E(q)$ (terms in $q^{+}$ but not in $q$), the \emph{Hallucinated
Expansion Term Rate} is
\begin{equation}\label{eq:hetr}
  \mathrm{HETR} = \frac{1}{|\mathcal{Q}|}\sum_{q\in\mathcal{Q}}
  \frac{\bigl|\{\,t\in E(q): \nexists\, d\in \mathrm{Rel}(q)\ \text{that supports } t\,\}\bigr|}{|E(q)|},
\end{equation}
where supports is decided by a high-precision verifier i.e., exact/near-lexical
match for entities and numbers, entailment for phrases. HETR is the fraction of
injected terms unsupported by any relevant document, a query-side analogue of hallucination score of HALoGEN. We report HETR alongside effectiveness, a
useful method lowers HETR without lowering nDCG@10.

\section{Experimental Studies}\vspace{-10pt}
We conduct a series of experiments to evaluate our proposed method, SAFE-QE investigating the following research questions.
\textbf{RQ1} Does filtering hallucinated terms cost retrieval effectiveness?. \textbf{RQ2} How many hallucinated terms does SAFE-QE actually keep out of the query relative to generative and corpus steered expansion?. \textbf{RQ3} Does the benefit concentrate on strong retrievers, the regime in which generative expansion is known to regress~\cite{weller2024generative}?. \textbf{RQ4} Does the faithfulness benefit survive a small generator, or does it merely track generator quality?.\\
\textbf{Datasets and Setups:} We employ test collections such as TREC Deep Learning 2019 (43 judged queries) and 2020 (54), and four BEIR
collections chosen to span domain shift and entity-heavy content, i.e. TREC-COVID (50), NFCorpus (323), FiQA-2018 (648) and SciFact (300).

We pair each collection with three first stage retrievers of increasing
strength, BM25 (sparse), SPLADE (learned sparse) and Contriever (dense). The three way split is what lets us separate expansion helps from `expansion helps weak retrievers', which differs from Weller et
al.~\cite{weller2024generative} draw and the one RQ3 turns on. Candidate terms are generated by four open, instruction-tuned LLMs, which are Qwen2.5-0.5B, Gemma-2-2B, Qwen2.5-7B and Gemma-2-9B. The reported results use the strongest generator, Gemma-2-9B, and the full generation is employed for RQ4. We employ Pyserini with prebuilt indexes for indexing and retrieval. Generation is greedy-free sampling with $N$ samples per query and $P$ perturbed prompts aforementioned in Section 3. We assess significance with paired $t$-tests over per-query scores, Bonferroni correction across baselines at $p<0.05$.\\
\\
\textbf{Baselines:} We employ the first stage retriever, no expansion on the raw query. Two generative baselines expand with free LLM text, Query2doc~\cite{wang2023query2doc} concatenates a generated pseudo document to the query, and HyDE \cite{gao2023precise} encodes a hypothetical document. For corpus steered, we employ two baselines condition generation on retrieved evidence, CSQE~\cite{lei2024corpus} extracts pivotal sentences from the top-$k$, and MILL~\cite{jia2024mill} cross verifies generated and retrieved
text. The generative pair isolates how much damage ungrounded generation does, the corpus steered pair is the strongest prior mitigation and an appropriate comparison for grounding term $\gamma$ of SAFE-QE.



\subsection{Metrics}
For effectiveness, we employ nDCG@10. 
Faithfulness employs HETR from Eq.~\ref{eq:hetr}. HETR is a rate, so it says nothing about how many terms a method injects in the first place. A method that adds two terms of which one is unsupported scores worse than one that adds eighty of which thirty are. We therefore also report the quantity a retriever actually sees, the \emph{hallucinated term load}
\begin{equation}
  \textrm{load} \;=\; \mathbb{E}_q\!\left[\,\textrm{HETR}(q)\cdot|E(q)|\,\right],
  \label{eq:load}
\end{equation}
the expected number of unsupported terms injected per query. Rate and load answer different questions and we report both throughout.


\subsection{Result}
\begin{table}[t]
\centering
\setlength{\tabcolsep}{2.8pt}
\caption{Paired per-query tests of SAFE-QE with each baseline, pooled over the four BEIR collections with the gemma-2-9b generator. $\Delta$ is SAFE-QE baseline. $p$ is a paired $t$-test with Bonferroni correction across baselines, $^{*}p{<}0.05$, $^{**}p{<}0.01$, $^{***}p{<}0.001$. On effectiveness SAFE-QE is statistically indistinguishable from every generative baseline, on faithfulness it is better than rest of the methods. The HETR of $0.440$ here is the BEIR-pooled per-query value; the all-collection macro-average appears in Table~\ref{tab:generator}.}
\label{tab:significance}
\begin{tabular}{l rrr r@{\hskip 10pt} rrr r}
\toprule
& \multicolumn{4}{c}{nDCG@10} & \multicolumn{4}{c}{HETR } \\
\cmidrule(lr){2-5}\cmidrule(lr){6-9}
Baseline & SAFE-QE & base & $\Delta$ & $p$ & SAFE-QE & base & $\Delta$ & $p$ \\
\midrule
No expansion & 0.404 & 0.410 & -0.006 & 0.152 & -- & -- & -- & -- \\
Query2doc & 0.403 & 0.411 & -0.008$^{*}$ & 0.021 & 0.440 & 0.582 & -0.142$^{***}$ & $<$0.001 \\
HyDE & 0.403 & 0.399 & +0.004 & 1.000 & 0.440 & 0.593 & -0.153$^{***}$ & $<$0.001 \\
CSQE & 0.404 & 0.407 & -0.004 & 0.683 & 0.369 & 0.416 & -0.047$^{***}$ & $<$0.001 \\
MILL & 0.404 & 0.400 & +0.004 & 0.562 & 0.369 & 0.435 & -0.066$^{***}$ & $<$0.001 \\
\bottomrule
\end{tabular}
\end{table}
\begin{table}[t]
\centering
\setlength{\tabcolsep}{2.3pt}
\caption{Evaluation across six collections and three retrievers, where generator is gemma-2-9b. The best nDCG@10 scores are in bold. \emph{n/a} represents the baseline's injected no terms at all and is slightly identical to no expansion, CSQE and MILL under SPLADE. These are excluded from every aggregate. $\dagger$ marks Contriever cells not included in this comparison; they are omitted from all aggregates. $\overline{\Delta}$ is the mean gain over no expansion a method covers.}
\label{tab:main}

\scalebox{0.8}{%
\begin{tabular}{llrrrrrr}
\toprule
Collection & Retriever & None & Query2doc & HyDE & CSQE & MILL & SAFE-QE \\
\midrule
DL’19 & BM25 & 0.5058 & \textbf{0.6324} & 0.5255 & 0.5611 & 0.5174 & 0.5876 \\
 & SPLADE & \textbf{0.7308} & 0.5848 & 0.6160 & \textit{n/a} & \textit{n/a} & 0.6950 \\
 & Contriever & 0.5812 & $\dagger$ & $\dagger$ & 0.5844 & $\dagger$ & \textbf{0.6381} \\
\addlinespace[1pt]
DL’20 & BM25 & 0.4796 & \textbf{0.6044} & 0.5597 & 0.5057 & 0.4622 & 0.5336 \\
 & SPLADE & \textbf{0.7197} & 0.6409 & 0.5936 & \textit{n/a} & \textit{n/a} & 0.6127 \\
 & Contriever & 0.5251 & \textbf{0.6731} & $\dagger$ & 0.5469 & $\dagger$ & 0.5775 \\
\addlinespace[1pt]
TREC-COVID & BM25 & 0.5947 & \textbf{0.7061} & 0.6430 & 0.6510 & 0.6825 & 0.6644 \\
 & SPLADE & \textbf{0.7270} & 0.6387 & 0.6574 & \textit{n/a} & \textit{n/a} & 0.6805 \\
 & Contriever & 0.5964 & \textbf{0.7321} & 0.7181 & 0.6740 & 0.6640 & $\dagger$ \\
\addlinespace[1pt]
NFCorpus & BM25 & 0.3218 & \textbf{0.3496} & 0.3301 & 0.3447 & 0.3305 & 0.3479 \\
 & SPLADE & 0.3473 & 0.3486 & \textbf{0.3556} & \textit{n/a} & \textit{n/a} & 0.3504 \\
 & Contriever & 0.3281 & \textbf{0.3372} & 0.3318 & 0.3307 & 0.3067 & 0.3221 \\
\addlinespace[1pt]
FiQA & BM25 & 0.2361 & \textbf{0.2583} & 0.2186 & 0.2386 & 0.2213 & 0.2401 \\
 & SPLADE & \textbf{0.3473} & 0.3080 & 0.3109 & \textit{n/a} & \textit{n/a} & 0.3218 \\
 & Contriever & \textbf{0.3293} & 0.3212 & 0.3073 & 0.2959 & 0.2832 & 0.3066 \\
\addlinespace[1pt]
SciFact & BM25 & 0.6789 & 0.7079 & \textbf{0.7095} & 0.6808 & 0.6907 & 0.6778 \\
 & SPLADE & \textbf{0.7036} & 0.6916 & 0.6904 & \textit{n/a} & \textit{n/a} & 0.6952 \\
 & Contriever & 0.6768 & \textbf{0.7054} & 0.6837 & 0.6765 & 0.6637 & 0.6808 \\
\addlinespace[1pt]
\midrule
\multicolumn{2}{l}{$\overline{\Delta}$ vs.\ no expansion} & -- & +0.0231 & -0.0045 & +0.0197 & +0.0075 & +0.0058 \\
\bottomrule
\end{tabular}%
}

\end{table}\vspace{-8pt}
We observe that the effectiveness result is a null one. SAFE-QE is statistically indistinguishable from HyDE, CSQE and MILL i.e., all $|\Delta| \le 0.004$, $p > 0.5$ after correction and within 0.008 of
Query2doc, where the gap reaches significance ($p = 0.021$) but is a tenth of a rank position at nDCG@10. Removing terms that no relevant document supports does not remove terms the retriever relies on. It can be observed via faithfulness as per RQ2 should be read against this. It yields at no measurable cost in ranking quality, which is the trade a practitioner is asked to make, which prior filtering work lacks. The justification for RQ1 is reported in Table~\ref{tab:main}, which shows that the effectiveness is not the price of faithfulness, and Table~\ref{tab:significance} reports the paired per-query tests, pooled over the four BEIR collections with the Gemma-2-9B generator.
\begin{table}[t]
\centering
\setlength{\tabcolsep}{4.6pt}
\caption{Faithfulness and query time cost on the collection, retriever, where generator is gemma-2-9b, macro-averaged over all six collections and three retrievers. HETR here is the micro-rate over injected terms, so it differs from the BEIR-pooled per-query value in Table~\ref{tab:significance}. $|E(q)|$ is the number of terms a method injects into the query. \emph{load} is the absolute number of those terms that no judged relevant document supports, HETR\,$\times\,|E(q)|$, and \emph{cut} is SAFE-QE's reduction in that load. Lower is better in the first three columns. SAFE-QE buys its faithfulness with generation cost, two sampled generations plus $P$ perturbed prompts.}
\label{tab:faithfulness}
\begin{tabular}{lrrrr rrr}
\toprule
& \multicolumn{4}{c}{Faithfulness} & \multicolumn{3}{c}{Query-time cost} \\
\cmidrule(lr){2-5}\cmidrule(lr){6-8}
Method & HETR & $|E(q)|$ & load & cut & s/query & tok/query & gen.\ calls \\
\midrule
Query2doc & 0.656 & 43.0 & 27.83 & 85\% & 4.06 & 78 & 1.00 \\
HyDE & 0.663 & 87.2 & 57.47 & 93\% & 1.10 & 231 & 0.19 \\
CSQE & 0.531 & 14.8 & 7.90 & 45\% & 5.23 & 758 & 1.00 \\
MILL & 0.546 & 13.8 & 7.52 & 43\% & 2.91 & 757 & 0.50 \\
SAFE-QE & \textbf{0.465} & 10.3 & \textbf{4.79} & -- & 10.97 & 1516 & 2.00 \\
\bottomrule
\end{tabular}
\end{table}
\begin{figure}[h!]
  \centering
  \includegraphics[width=\linewidth]{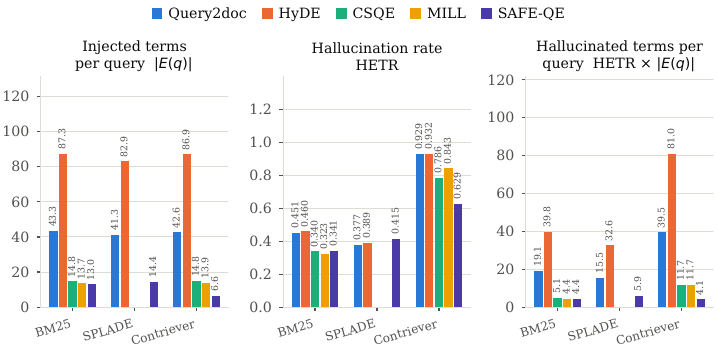}
  \caption{Results for decomposing what a retriever actually receives, per retriever, over the collections. On the left, how many terms a
  method injects. In middle, what fraction of them no judged relevant document supports (HETR). On right, their product, the hallucinated-term load. The advantage of SAFE-QE is largest on the dense retriever, where free generation is least grounded.}
  \label{fig:load}
\end{figure}

\begin{table}[t]
\centering
\setlength{\tabcolsep}{4.6pt}
\caption{HETR across all six collections and three retrievers, and generator is gemma-2-9b. HETR is retriever-independent for Query2doc and HyDE, whose injected set does not depend on the first stage, and retriever-dependent for the grounded methods, whose selection is computed against the retriever's own pseudo-relevant set. \emph{n/a} and $\dagger$ are as in Table~\ref{tab:main}.}
\label{tab:hetrgrid}

\scalebox{0.8}{%
\begin{tabular}{llrrrrr}
\toprule
Collection & Retriever & Query2doc & HyDE & CSQE & MILL & SAFE-QE \\
\midrule
DL’19 & BM25 & 0.194 & 0.206 & 0.042 & 0.065 & \textbf{0.041} \\
 & SPLADE & \textbf{0.194} & 0.206 & \textit{n/a} & \textit{n/a} & 0.310 \\
 & Contriever & $\dagger$ & $\dagger$ & 0.274 & $\dagger$ & \textbf{0.125} \\
\addlinespace[1pt]
DL’20 & BM25 & 0.262 & 0.288 & \textbf{0.077} & 0.106 & 0.083 \\
 & SPLADE & \textbf{0.262} & 0.288 & \textit{n/a} & \textit{n/a} & 0.370 \\
 & Contriever & 0.747 & $\dagger$ & 0.347 & $\dagger$ & \textbf{0.197} \\
\addlinespace[1pt]
TREC-COVID & BM25 & 0.033 & 0.038 & \textbf{0.003} & 0.011 & 0.003 \\
 & SPLADE & \textbf{0.033} & 0.038 & \textit{n/a} & \textit{n/a} & 0.065 \\
 & Contriever & 0.658 & 0.670 & \textbf{0.096} & 0.349 & $\dagger$ \\
\addlinespace[1pt]
NFCorpus & BM25 & 0.481 & 0.481 & \textbf{0.265} & 0.353 & 0.282 \\
 & SPLADE & 0.481 & 0.481 & \textit{n/a} & \textit{n/a} & \textbf{0.464} \\
 & Contriever & 0.910 & 0.911 & 0.705 & 0.796 & \textbf{0.522} \\
\addlinespace[1pt]
FiQA & BM25 & 0.619 & 0.633 & 0.527 & \textbf{0.455} & 0.521 \\
 & SPLADE & \textbf{0.619} & 0.633 & \textit{n/a} & \textit{n/a} & 0.626 \\
 & Contriever & 0.928 & 0.934 & 0.814 & 0.836 & \textbf{0.645} \\
\addlinespace[1pt]
SciFact & BM25 & 0.673 & 0.690 & 0.566 & \textbf{0.473} & 0.557 \\
 & SPLADE & 0.673 & 0.690 & \textit{n/a} & \textit{n/a} & \textbf{0.657} \\
 & Contriever & 0.950 & 0.952 & 0.840 & 0.896 & \textbf{0.721} \\
\addlinespace[1pt]
\bottomrule
\end{tabular}%
}

\end{table}
We can observe that in Table~\ref{tab:faithfulness}, SAFE-QE has both the lowest hallucination rate 0.465 as compared to Query2doc (0.656) and HyDE (0.663), and much the smallest expansion footprint 10.3 terms per query against 43.0 and 87.2. Also, SAFE-QE puts 4.30 unsupported terms into the average query as compared to 27.83 for Query2doc and 57.47 for HyDE, a reduction of 85\% and 93\%. Against the corpus steered baselines, which selects a small grounded set and are therefore the hard comparison, the reduction is 45\% over CSQE and
43\% over MILL.

\begin{figure}[h!]
  \centering
  \includegraphics[width=\linewidth]{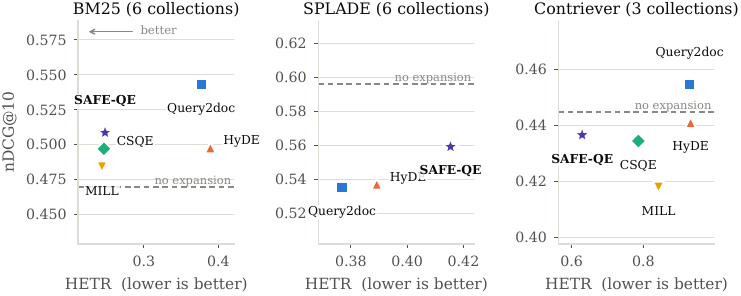}
  \caption{The faithfulness effectiveness plane, per retriever, macro averaged over the collections. Up and to the left is better. SAFE-QE sits on the frontier under BM25 and Contriever. Under SPLADE its
  rate is higher than the generative baselines' even though its
  load is a third of theirs (Fig.~\ref{fig:load}), because grounding is
  computed against SPLADE's own pseudo-relevant set and that set admits more candidates.}
  \label{fig:pareto}
\end{figure}

We report HETR for every collection and retriever in Table~\ref{tab:hetrgrid}. The BEIR restriction belongs to the significance analysis, where the per-query artefacts live. HETR is defined per query from the injected set and the judged relevant documents. For the individual collections, we report it via Fig.~\ref{fig:pareto}, which shows SAFE-QE has the lowest HETR on all five comparable Contriever, where free generation is least grounded. On BM25 it is lowest on one of six with CSQE and MILL keeping a marginally cleaner set from a comparable footprint, and on SPLADE it is lowest on two of six, the retriever-dependence discussed above. Its advantage in that last regime is in the footprint.
We report pooled per-query over BEIR in Table~\ref{tab:significance}, every HETR reduction is significant at $p<0.001$, -25.9\% relative against HyDE, -24\% against Query2doc, -15.1\% against MILL and -11.2\% against CSQE. If we recall RQ1 together with this, which  justifies that comparable ranking quality on a significantly more faithful expansion.

Figure~\ref{fig:pareto} shows the trade-off and also its one
exception. The grounding signal $\gamma$ of SAFE-QE is computed against the retriever's own top-$k$, so unlike Query2doc and HyDE, whose injected set, and therefore whose HETR is retriever-independent. SAFE-QE's selection moves with the first stage. Under SPLADE the pseudo-relevant set both passes more candidates through the grounding threshold 14.4 terms per query against 12.3 under BM25, and grounds them less reliably, and the rate rises with it from 0.248 to 0.415. The load still falls because the footprint is a third of Query2doc's, but the rate does not, and we report this to inform that the grounding threshold $\tau_g$ should be calibrated per retriever instead of fixed, which we leave to future work.

Under Contriever, on the three collections including five methods, every method is within $\pm 0.03$ of no expansion and the differences are not significant. Dense retrieval is the regime, where this separates the methods least on effectiveness, and can be observed from Fig.~\ref{fig:load} separates them most on faithfulness.
\begin{table}[t]
\centering
\setlength{\tabcolsep}{4.0pt}
\caption{nDCG@10 and HETR as the expansion LLM grows from 0.5B to 9B parameters, macro-averaged over all collections and retrievers. SAFE-QE's faithfulness advantage is present at every scale and without dependence on a large generator; the 9B HETR ($0.375$) therefore differs from the BEIR-pooled $0.440$ in Table~\ref{tab:significance}.}
\label{tab:generator}
\begin{tabular}{l rrrr r rrrr}
\toprule
& \multicolumn{4}{c}{nDCG@10} & & \multicolumn{4}{c}{HETR} \\
\cmidrule(lr){2-5}\cmidrule(lr){7-10}
Method & Qwen & Gemma & Qwen & Gemma & & Qwen & Gemma & Qwen & Gemma \\
& 0.5B & 2B & 7B & 9B & & 0.5B & 2B & 7B & 9B \\
\midrule
HyDE & 0.3951 & 0.4641 & 0.4697 & 0.4788 & & 0.640 & 0.597 & 0.590 & 0.570 \\
MILL & 0.4414 & 0.4535 & 0.4584 & 0.4620 & & 0.480 & 0.447 & 0.420 & 0.443 \\
SAFE-QE & 0.4648 & 0.4784 & 0.4750 & 0.4845 & & 0.398 & 0.386 & 0.344 & 0.375 \\
\bottomrule
\end{tabular}
\end{table}
We observe that SAFE-QE is better than both baselines reported in Table~\ref{tab:generator}. First, the effectiveness of SAFE-QE is far flatter in generator size than HyDE from 0.5B to 9B. 
The faithfulness benefit is a property of the filter and is irrespective
of the generator behind, because it is the cheap generator
that a production expansion pipeline can afford.

\section{Conclusion}
Our proposed approach follows the subsequence-association account of hallucination~\cite{sun2026and} and the consistency breadth stability characterisation of language generation~\cite{kalavasis2026characterizations} to a
long standing IR task. Our method, SAFE-QE is a retriever agnostic method that traces and downweights hallucinatory expansion terms and selects a set that is broad but consistent and stable across resampling. Alongside it, we introduced HETR, a query side hallucination metric for expansion, and argued that a rate must be together with the expansion footprint it applies to the hallucinated-term load is what a retriever actually receives. Across six collections and three retrievers, SAFE-QE matches generative and corpus-steered expansion on nDCG@10 while injecting $85\%$ fewer unsupported terms than Query2doc and $93\%$ fewer than HyDE, and $43$-$45\%$ fewer than the corpus steered baselines that represent the strongest prior mitigation.
The benefit is produced by the filter instead of a generator behind it, with a 0.5B generator SAFE-QE is already more faithful than HyDE with a 9B one. On a strong learned sparse retriever, where every expansion method regresses, SAFE-QE halves the regression. The grounding threshold $\tau_g$ is fixed and should be calibrated per retriever, since the SPLADE result shows the pseudo-relevant set, and setting how many
candidates survive. The component ablation is outstanding and is what would turn the design argument into a measured one. More broadly, our formulation extends to generative retrieval and to learned sparse expansion, and the finite sample behaviour of the breadth-consistency trade-off remains open. We use the asymptotic characterisation as a design vocabulary, and a finite
analysis of when a budget-$B$ grounded subset is broad enough would put that borrowing on firmer ground. We hope the bridge encourages further mechanism grounded IR methods.
breadth-consistency trade-off.

%
%
\bibliographystyle{splncs04}
\bibliography{refs}
\end{document}